\documentclass[acmlarge]{acmart}
\usepackage{graphicx,kantlipsum}
\usepackage{float}

\AtBeginDocument{%
  \providecommand\BibTeX{{%
    \normalfont B\kern-0.5em{\scshape i\kern-0.25em b}\kern-0.8em\TeX}}}
\usepackage{algorithm2e}

\begin{document}

\title{Profiling minisat based on user defined execution time - AccEx}
\author{Shubhendra Pal Singhal}
\authornotemark[1]
\email{106116088@nitt.edu}

\affiliation{%
  \institution{ \\ Department of Computer Science and Engineering\\National Institute of Technology, Tiruchirappalli\\}
}
\author{Sandeep Gupta}
\authornotemark[2]
\email{sandeep@usc.edu}

\author{Pierluigi Nuzzo}
\authornotemark[3]
\email{nuzzo@usc.edu}

\affiliation{%
  \institution{ \\ EEB, Viterbi School of Engineering\\University of Southern California\\}
}

\begin{abstract}
This project focuses on the explanation of the architecture of profilers particularly gprof and how to profile a program according to the user defined input of execution time . Gprof is a profiler available open source in the package of binutils. Gprof records the flow of the program including the callee and caller information and their respective execution time. This information is represented in the form of a call graph which is explained in more detail in section 4. Profilers at the time of execution creates a call graph file which indicates the full flow of the program including the individual execution time as well. This project aims at providing a better understanding of the data structure used to store the information and how is a profiler(gprof) actually using this data structure to give user a readable format. The next section of this project solves one of the limitation of gprof i.e. edit the time of block of code without understanding the call graph. Any changes in the execution time of a particular block of code would affect the total execution time. So if we edit the gprof in such a way that its consistent and platform independent, then it can yield various results like testing execution time after parallelism, before even designing it by replacing the values with theoretical/emulated ones and see if the total execution time is getting reduced by a desired number or not? Gprof edit can help us figure out that what section of code can be parallelized or which part of code is taking the most time and which call or part can be changed to reduce the execution time\cite{6}. The last section of the project walks through the application of gprof in minisat and how gprof helps in the hardware acceleration in minisat by suggesting which part to be parallelised and how does it affect the total percentage. AccEx tool is made available open source by project partners \cite{10}.  
\end{abstract}
\acmDOI{}
\renewcommand\footnotetextcopyrightpermission[1]{} 
\pagestyle{plain} 
\settopmatter{printacmref=false}
\setcopyright{none}
\renewcommand\footnotetextcopyrightpermission[1]{}
\pagestyle{plain}
\maketitle

\section{Introduction}
\subsection{SAT}
A problem of determining whether the variables of a given Boolean formula can be consistently replaced by the values TRUE or FALSE in such a way that the formula evaluates to TRUE.\newline
Nowadays, SAT is finding a lot of applications in the field of (Hardware) Equivalence Checking, Asynchronous circuit synthesis (IBM), Software-Verification, Cryptanalysis\cite{13,14}.
\subsection{Minisat}
MiniSat is a minimalistic, open-source SAT solver\cite{12}.
\begin{itemize}
\item[1] Easy to modify 
MiniSat is small and well-documented, and also well-designed, making it an ideal starting point for adapting SAT based techniques to domain specific problems.\newline
\item[2] Highly efficient and Designed for integration. MiniSat supports incremental SAT and has mechanisms for adding non-clausal constraints\cite{11}. By virtue of being easy to modify, it is a good choice for integrating as a backend to another tool, such as a model checker or a more generic constraint solver.
\end{itemize}
\subsection{Profiling programs}
Profiling is a dynamic program analysis\cite{2} that measures the time complexity of every instruction in a program. The information is stored by using -pg flag in gcc\cite{5}. This flag when enabled switches debugging and profiling on during the compilation of the program. This produces an executable file. This executable file when run produces a file "gmon.out" which contains the call graph information of the program. This is a binary file and is not readable by a user. So, gprof uses the symbol table and histograms and records every instance in a histogram. Histogram contains the address and the execution time of that instance. The symbol table consists of the function name and its corresponding assigned index. This information is traced by the profiler to print the call graph of the program in a readable format\cite{5}. The profiler uses the approach of traversing the path from root to the last child (in that arc) in the graph which helps to figure out that what calls have been made to whom and what was the time taken inside a parent and the child separately. This information is stored in the form of arcs making parent as the start pointer. The information is then printed in such a way that there are two ways to discern the data flow\cite{1} : \newline
1.) Flat Profile : The flat profile shows the total amount of time your program spent executing each function. The functions are sorted by first by decreasing run-time spent in them, then by decreasing number of calls, then alphabetically by name. Generally, sample count is 0.01 sec indicating that the number 'x' units in the profile represents (100 * x) seconds of the execution.\cite{1}\newline
2.) Call Graph Profile : The call graph shows how much time was spent in each function and its children. In each entry, the primary line is the one that starts with an index number in square brackets. The end of this line says which function the entry is for. The preceding lines in the entry describe the callers of this function and the following lines describe its subroutines(children).\cite{1}
\subsection{Insight of solution}
SAT applications seen above, involve millions of variables to be solved (like in circuit verification). This issue can be addressed and solved by accelerating SAT Solver either by hardware or software. Software optimizations and parallelism are being worked on, but simultaneously question arises : Why not try with hardware?\newline
There is a possibility of acceleration of SAT Solvers by a (hardware) chip. Instead of designing a hardware chip and testing it whether it really works, we emulated the results of hardware and fed these inputs to SAT solver and analysed the reduction in total execution time. This was achieved by using open source software AccEx(variation of Gprof created by authors of project). It analyses the flow of program and execution time at every call or instruction of a program.
\section{GPROF}
\subsection{What is GPROF?}
Gprof is a profiling program which collects and arranges statistics on your programs.Basically, it looks into each of your functions and inserts code at the head and tail of each one to collect timing information 
\subsection{Technical terms related to Gprof}
1.) Call Graph : It is a graph which describes the calling pattern of the functions and analyses the execution time of every call even if it involves the same parent and child.\cite{1} \newline
2.) Self time : The total execution time of the function excluding the time taken by its subroutines(children) is called self time.\cite{1}\newline
3.) Child time : The time taken by the children of a function i.e. the difference between the time instant when it returns back to caller and the time instant the callee is called by the caller.\cite{1}\newline
4.) Arc : Arc denoted the path from the root to leaf indicating the calls made by the root and then subsequent callers(children of root) till the point the last child returns back to its caller\cite{1}.\newline
5.) Histogram : It is a data structure which consists of a low base address and high end address which represents the storage of the callee and its corresponding caller. The bin\_count represents the total execution time taken by an instance. It can be used to describe the self time and child time. If the function represents the child of a function, then the corresponding bin\_count will represent the self time of child. For example if func1() calls func2(), then the arc func1()----> func2() will represent the time taken by func2() alone. The func1() time will be calculated by the arc main()--->func1(). So like this there will be one root function which will not have any parent. This function is named as spontaneous and the time recorded for this function is stored in the arc representing spontaneous as the tag. The hist time represents the self time taken by the function and the "child\_time" represents the time taken by its children.\newline
6.) Propagate time : The self time taken by the function in an arc is called prop.self time. The prop.child time represents the time taken by the children of that function keeping in mind that if already traversed arcs includes the child, then it needs to be included. \newline
7.) Bin-count : In this project, the bin\_count is referred to as the time corresponding to the bin of histogram. The code has a totally different interpretation of bin\_count which is different from what is used in this project. It is not the count of the bin in histogram but its the time which is obtained after scaling using the sample count as a metric. \newline
Every profile shows the average time per call\cite{3}.
\subsection{Skeletal structure of Gprof}
Gprof uses the gmon.out and executable file to extract the execution time in histogram (specified in hist.c). Call graph is constructed from gmon.out by "call\_graph.c" These bins then are collectively merged to get the total and self time of a function in hist.c. This is used to create flat profile. Using this "cg\_arcs.h", gprof constructs the arc from the parent to the leaf of the graph. This arc is then printed in an order specified by call graph profile and the execution time is taken from flat profile. Supposedly, a function took 30sec averagely with 3 calls, the cal graph will show 10sec/call\cite{8} in the call graph profile, thus extracting this average time from flat profile, shown in file cg\_print.c". Apart from these files, rest are all utility or library based files that support these files - hist.c, cg\_arcs.h, cg\_print.c\cite{1}.
\subsection{Illustration of how call graph is formed using sample program}
\begin{figure}
    \includegraphics[scale = 0.5]{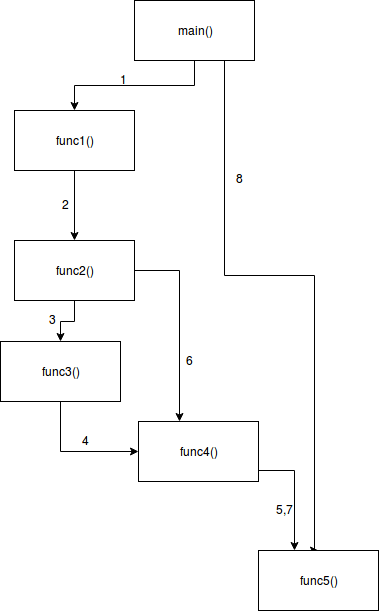}
    \caption{Illustration of Call Graph}
\end{figure}{}
\begin{figure}
    \includegraphics[scale = 0.6]{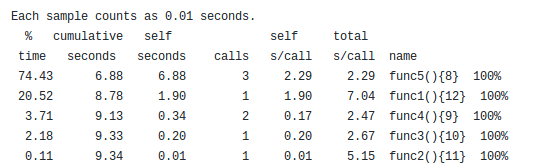}
    \caption{Flat Profile}
\end{figure}{}
\begin{figure}
    \includegraphics[scale = 0.6]{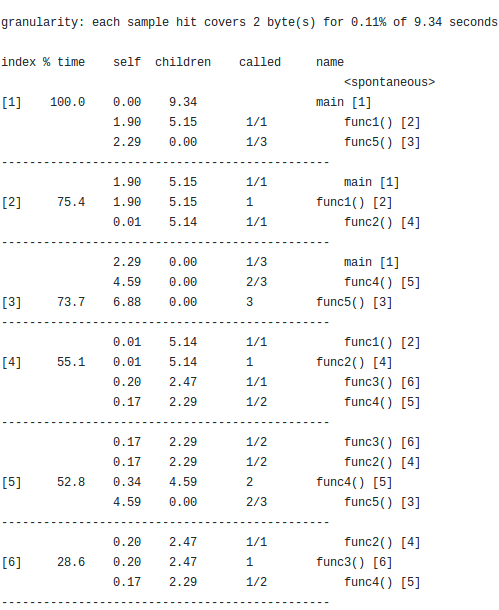}
    \caption{Call Graph Profile}
\end{figure}{}
Fig1. represents the call graph of the program. The calculation of time is always a bottom top approach as the time can be calculated when the call returns from its child. So the call-graph profile is topologically sorted and if there any recursive calls or cycles then that cycle is considered as one strongly connected component and the further topological sorting is done considering it as one node. The call graph profile thus starts with the arc representing func3(), func4() and func5(). Calculation of this arc means that first calculation of func3() can be done as both  func4() and func5() are getting called in the subsequent execution. Now, the func4() is called once and func5() is called once. So, when we proceed to index 5, we need to include two calls of func4()--> func5() as there are 2 calls of similar kind. The average of time taken by two calls is taken. Similarly, the full call graph is constructed as shown in Fig 3.\newline
The Fig 2. shows the percent of time spent and their respective total time. Cumulative seconds helps to determine the total time taken i.e. total execution time of the program.

\section{Prerequisites for running AccEx}
\subsection{Installation of GPROF in LINUX environment}
\textbf{sudo apt-get install binutils} is the command which will install the package binutils. Gprof is a sub-directory inside this directory\cite{4}.

\section{Discovery of AccEx}
\subsubsection{Limitation of Gprof}
One of the limitation of gprof\cite{3} that the arc time in call graph will always be average\cite{8}. Even if one of the arc takes very less amount of time in one case and the same arc takes more time in another case, still the time displayed will be average of both and not the actual one. This involves the need to understand the call graph before editing the block of code according to user defined input. But big programs like SAT has a lot of functions and thus, it makes a bit difficult to trace all functions and then edit the time for that block of code. Thus, we have to develop a tool which can analyze the pattern of how is the total execution time getting affected without understanding the call graph.
\subsection{Changing the printing pattern of gprof}
In fig3., suppose we want to change the self time taken by func5(). An "if" statement would suffice it stating that if the index is equal to 5, then change the time. At this point of time, there is no problem, but if we move a step up and analyze func3() time, then we cannot trace that func3() called func5() effectively i.e. though func3() called func4(), but the execution time of func3() will change as the time taken by func4() is also changed. So, this particular change can be propagated to its immediate next parent i.e. func4(), but not to the whole graph as we don't have any means to store every function calling children. Visiting every child when a parent is encountered will take a huge amount of time and the time also will not be stored anywhere meaning that if we want to retrieve any arc time that is getting affected by this change, then we cannot get it as we are just printing and not changing the values of assigned data structure i.e. histogram \cite{2}.

\subsection{Editing data structure of gprof}
Suppose, the part of program can be replaced by some coding paradigm\cite{2} which helps to reduce the execution time of that particular block of code, then we have to replace that time and see that how execution pattern gets affected. This requires a change in the record of histogram corresponding to the block. Identifying the block can be done by 2 ways : \newline
1.) Address - The address of the block can be found via printing and the corresponding block can then be changed. But there is one flaw here and that is if we run the program next time, we are not sure whether the program will load in the exact same address or not. This is what we call as platform dependency where every instance is changed.\newline
2.) Index - The above idea seems to be a bit infeasible. So in this situation we analyze that what is not platform dependent. Address, time etc. are all machine dependent, but the call graph is not. So we assign the index to every call and then change the time for those calls(corresponding to the block we need to change). Index is assigned to every call and the respective id can be printed so that the user can identify the call by noticing its parent and respective child. The time corresponding to this id can be changed in the code. This change will be propagated and the final execution time will also get changed by the same amount.\newline
\subsection{Implementation of AccEx}
The histogram consists of bin\_count which is used to indicate the time taken in execution.\newline
The algorithm described below is edited in the file "hist.c" which contains the histogram data structure \cite{2} that is built from "gmon.out" file. The function "static void
hist\_assign\_samples\_1 (histogram *r)" consists of assigning the address and time to a histogram, so if we change the time here corresponding to the id that we want to change, the the entire pattern can be generated. The id that needs to be changed is known by printing it for every call initially and then figuring out which id corresponds to which call. A table of id and corresponding call is printed which helps to identify which id needs to be changed by looking at which call has to be modified. Ids are not machine dependent which makes it a feasible way of changing time of a specific block.  
\begin{algorithm}[H]
\caption{Algorithm for changing execution time of a certain part of code}
\SetAlgoLined
\hrule
\vspace{6pt}
Assume min to be lower limit and max as upper limit of block that needs to be replaced. c be the changed value. In case of different values for every id, it needs to be specified in the form of an array of values. \\
 $id = 0$, $count = 0$, $total\_bin = 0$ \\
 \If{$id>=min-1$ and $id <= max-1$}{ 
        $total\_bin$ += $bin\_count$ \\ 
        $bin\_count$ = $c$ }
      $count = count+1$ \\
    $count\_time$ = $bin\_count$
   \hrule
\vspace{6pt}
\end{algorithm}
\section{Results obtained}
\subsection{Compilation and results obtained for sample problem}
There are two types of gprofs that are required :\newline
1.) Gprof : hist.c is edited to print the ids and the corresponding calls.
2.) AccEx : After identifying the call that needs to be changed, the algorithm is included in hist.c \newline
After editing hist.c, gprof has to be compiled \cite{5} and the changes needs to be incorporated, so we use : \newline
1.) \textbf{make} (rectify the errors if any) \newline
2.) \textbf{make install} \newline
This will create version1(GPROF) and version2(AccEx). Gprof is used on the executable file and "gmon.out" as : \newline
\textbf{<directory-name>/gprof -d <.exe file> gmon.out > <file-name>} \newline
e.g. The program shown in Fig3.\cite{7} is executed and the exe file is test. The gprof directory is /home/. So, the command is :
/home/gprof test gmon.out > a.txt\newline
This will save the output in a.txt file. So we need to create two such text files corresponding to gprof and AccEx.

\subsubsection{Analysis of results obtained} 
\begin{figure}
    \includegraphics[scale = 0.4]{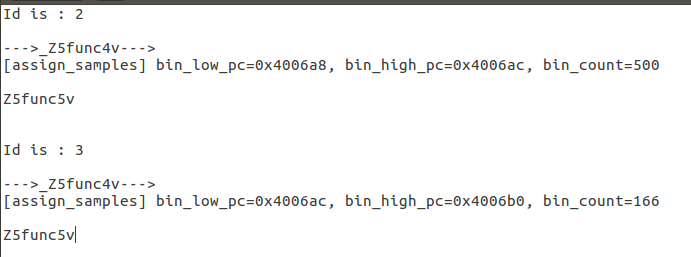}
    \caption{Bin id print for GPROF}
\end{figure}{}
\begin{figure}
    \includegraphics[scale = 0.4]{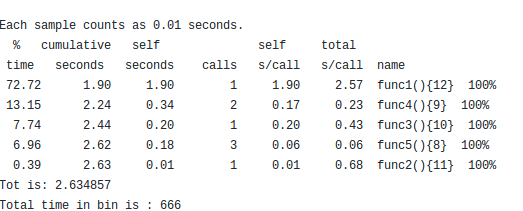}
    \caption{Flat Profile version2 - AccEx}
\end{figure}{}
\begin{figure}
    \includegraphics[scale = 0.4]{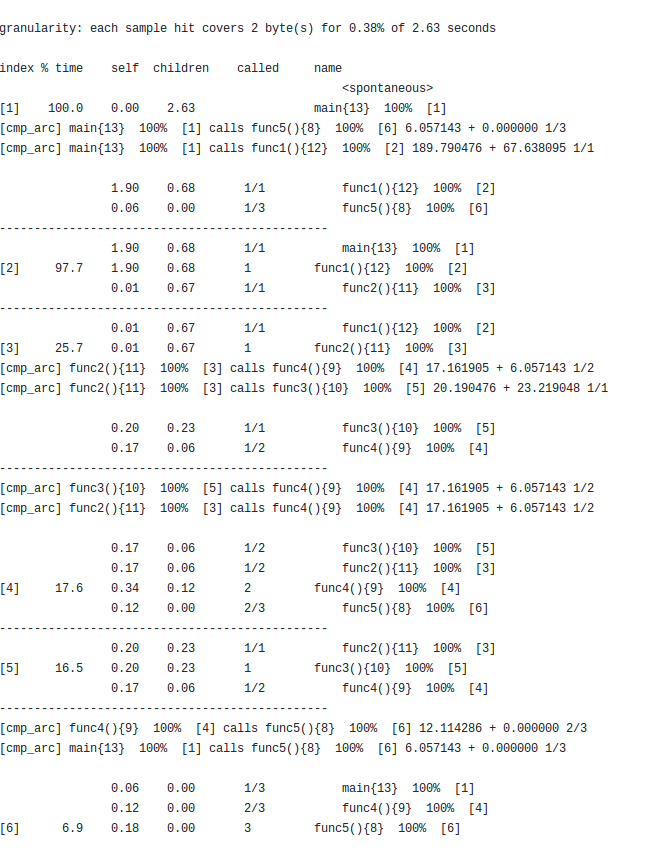}
    \caption{Call Graph Profile version2 - AccEx}
\end{figure}{}
Suppose, we want to change the time of arc func4()---> func5() to 0.01 seconds, then this particular bin id checked is 2 and 3 as shown in fig4. The func5() time needs to be changed indicating that func4() should be the parent and func5() must be the child in the representation and only bin ids 2 and 3 correspond to this arc. Suppose we change both of them to 1 which implies that the total time taken by func5() must reduce from  6.88 to \\ 0.16( start-->func5() ) + 0.01*2(1 sec*sample-size = 1*0.01 = 0.01, and there are two such calls) which equals 0.18.\newline
Difference 6.88 and 0.18 = 6.7 i.e. the total execution time should reduce by 6.7, i.e. from 9.34 to 9.34-6.7 = 2.64 which is close to the actual result obtained in fig 5.
Also when we analyze the call graph profile, then we notice that func5() takes just 0.06 seconds per call which is correct because 0.18/3 = 0.06 in fig5 and fig6.
\begin{figure}
    \includegraphics[scale = 0.6]{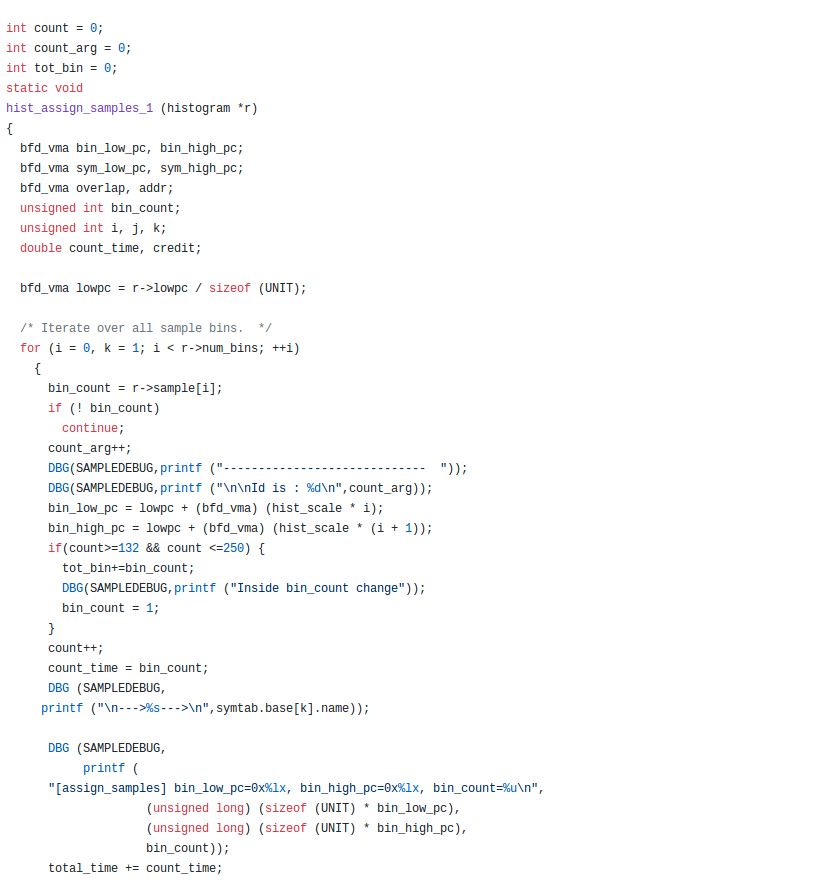}
    \caption{Editing hist.c for user defined execution time}%
    {\small Count corresponds to id of the bin. The id corresponding to the change is written in if condition e.g. here it is 133-251. See the total bin count which is being replaced and getting stored in tot\_bin. User defined input is bin\_count = 1.}
\end{figure}{}

\subsection{Compilation and analysis of the result obtained for sample problem for Minisat }
Using gprof, we concluded that propagate() function took almost 70-80\% time of minisat and specifically the arc analyze()---> propagate() as suggested in fig2.
This arc took 70-90\% of the time in propagate function. Gprof analysis  is shown for standard sat benchmark \cite{9} in fig9. 
\begin{figure}
    \includegraphics[scale = 0.35]{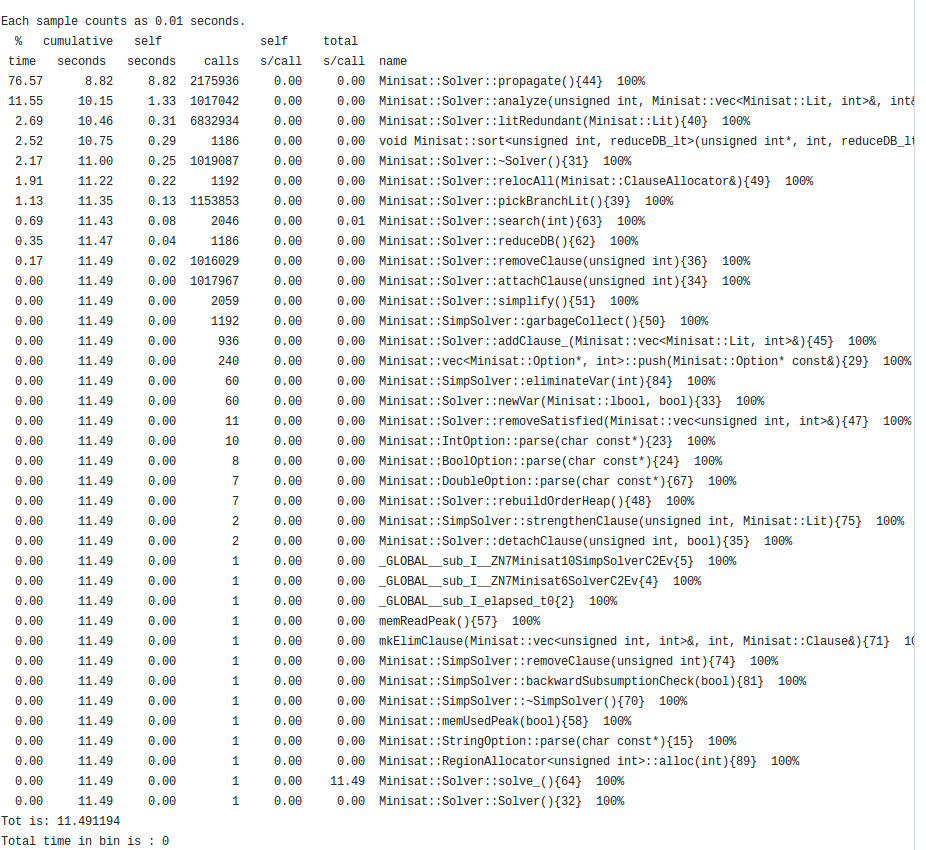}
    \caption{Call Graph Profile (GPROF) for minisat}
\end{figure}{}
The total time taken by minisat is 11 sec out of which propagate took around 8.8 sec which accounts for 76\% of total time implying that propagate() accounts for such a huge execution time.
This analysis led us to emulate the propagate() function and replace the bin\_count with a virtual value (indicated in fig10) and the respective results were obtained which indicates a significant improvement of 36\% in the execution time of minisat provided there is a way to parallelise propagate() function. This suggests that the hardware which can support propagate() function parallelism needs to be designed. 
\begin{figure}
    \includegraphics[scale = 0.35]{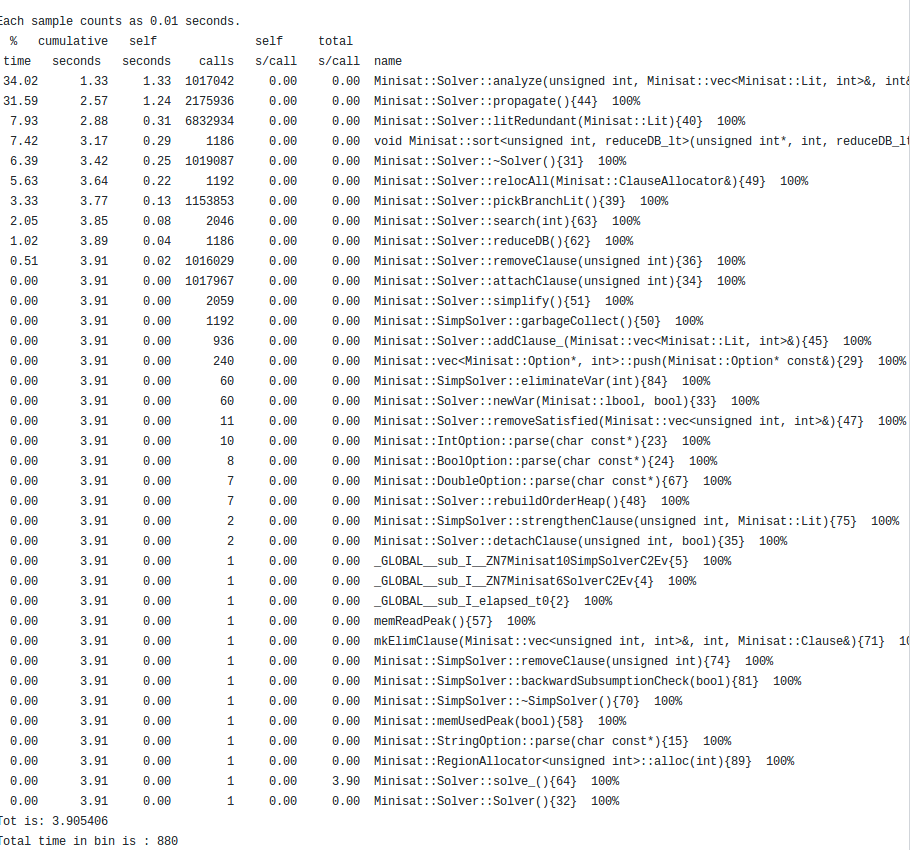}
    \caption{Call Graph Profile version2 for minisat}
\end{figure}{}
\subsection{Results of AccEx}
Now, we know that if function which needs to be parallelised is known, then its respective bin\_count can be changed using Algo 1. But the real question is which pfunction should actually be parallelised? 
\subsubsection{Which function should we actually parallelize - Minisat ?}
We analyse the execution time of every function in the program and the one with the highest execution ti me needs to be parallelised. This is obtained from the analysis of gprof. As we can see from Fig1, we can notice that propagate() function takes the most amount of time, and thus this function is the oen which needs to be parallelised.
\subsubsection{How does propagate() affect total execution time - Minisat?}
Supposedly, we reduce the propagate time by 10\%, what is the \% reduction of total execution time.
\newline
\begin{figure}
    \includegraphics[scale = 0.3]{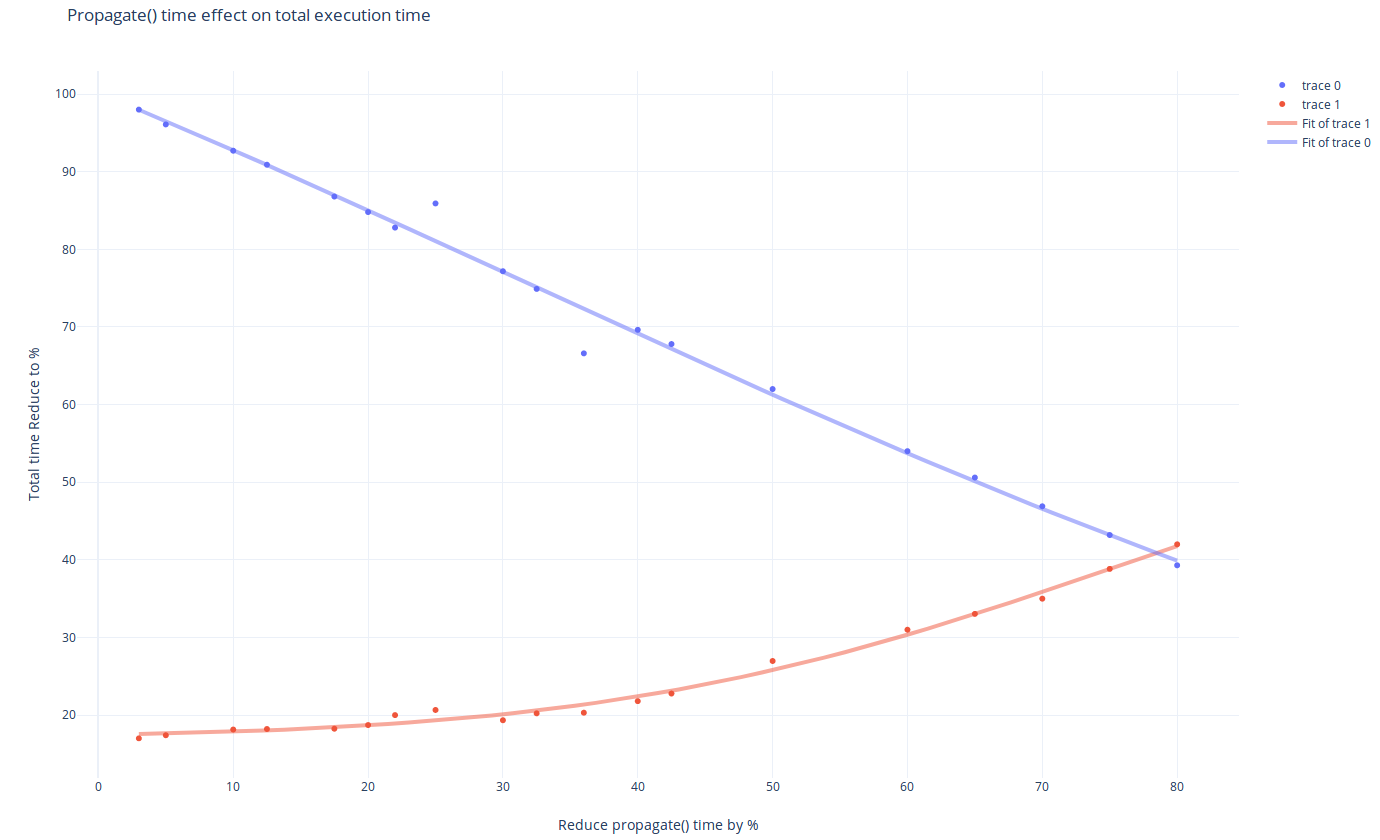}
    \caption{Effect of reduction in propagate time on total time}
\end{figure}{}
We analyse the result of AccEx tool by parallelising propagate() function and plotting a graph between \% reduction of propagate() vs \% reduction of total time due to propagate() as shown in fig11. Blue line is represented by Gaussian Curve which represents the reduction in total time when propagate() time is reduced to a certain percentage. While, we reduce propagate() time, the \% wise time of other functions such as analyse() and litRedundant() starts increasing which shoots up as in this case analyse() exponentially grew from 17.6\% to 42\%, litRedundant() grew from 2.1\% to 5.18\%, thus giving us the threshold that we should reduce propagate() time till its comparable with other functions(intersection point of graph).
\section{Conclusion}
Changing the time of a particular block of code could be done in several ways but considering the factor that code should be portable\cite{6}, reference to a call by its id turned out to be a suitable method. Gprof printing pattern cannot be changed as the calls cannot be traced back as seen in the section 4.1. Adding an extra attribute to a data structure is a small change that can profile a program according to the user needs. This strategy can be used to test that how much does emulated time of a block of code affects the total execution time. If its an expected number, then the hardware design\cite{6} for the same can be implemented on the chip instead of first designing it and then finding it to be futile. This can improve the hardware design process and also can help see whether any kind of parallelism is helping in execution as we saw in minisat.  \newline
\section{Acknowledgement}
We have taken efforts in this project.      However, it would not have been possible
without the kind support and help of many individuals and organizations.      We
would like to extend our sincere thanks to all of them. We are highly indebted to Prof. Dr. Pierluigi Nuzzo and Prof. Dr. Sandeep Gupta for their guidance and constant supervision as well as for providing necessary information regarding the project also for their support in completing the project. Our thanks and appreciations also go to all the team members in developing
the project and people who have willingly helped us out with their abilities.

\bibliographystyle{ACM-Reference-Format}
\bibliography{sample-base}

\end{document}